\documentclass[12pt]{article}
\usepackage{graphicx}
\usepackage{axodraw}
\usepackage{amstext}
\usepackage[small]{caption}

\newcommand{\p}{\mathbf{k}_{\perp}^2}
\renewcommand{\Im}{\mathop{\mathrm{Im}}\nolimits}
\begin{document}
\vspace{4cm}
\begin{center}\textbf{Calculation of the cross section for the pion}\\
\textbf{diffractive dissociation into two jets}\\[0.5cm]

V.L.~Chernyak and A.G.~Grozin\\[0.2cm]
Budker Institute of Nuclear Physics,\\
630090 Novosibirsk, Russia\\[1cm]
\textbf{Abstract}
\end{center}

Contribution of the gluon distribution in the nucleon to the cross section 
for the pion diffractive dissociation into two jets has been calculated 
recently. In addition, we present here calculation of
contribution of the quark distributions to this process.

It is expected usually that the quark contributions are much 
smaller than the gluon one at high energies.
We show that, under the conditions of the E791 experiment, the quark
contributions are as significant as the gluon one, so that the total cross
section exceeds several times those with only gluon contribution.

Nevertheless, the distribution of jets in longitudinal momentum
fractions does not change significantly and remains qualitatively the same.
We compare our results with the data from the E791 experiment.
\newpage

\textbf{1.}\hspace{1cm} The E791 experiment at Fermilab~\cite{Aitala}
has measured recently the cross section of the hard diffractive
dissociation of the pion into two jets. In particular, the
distribution of the total pion longitudinal momentum into fractions $y_1$ and
$y_2\equiv1-y_1$, between jets has been measured. This distribution
is potentially of great interest because its form depends on the
pion wave function profile. So, its measurement can be used
to obtain information about properties of the nonperturbative pion wave 
function $\phi_{\pi}(x_1)$ which describes the distribution of
quarks inside the pion in the longitudinal momentum fractions 
$x_1$ and $x_2\equiv1-x_1$.
  
This cross section has been calculated in the papers~\cite{Ch0}
and~\cite{Braun}\footnote{In comparison with their original calculation,
the authors of~\cite{Braun} 
have found recently an additional missed contribution, so that their revised 
analytic results for the scattering amplitude agree now with those obtained 
in \cite{Ch0}.},
in the approximation when only the contribution of the gluon
distribution in the nucleon has been accounted for, while those of the
quark distributions were neglected. The purpose of this paper is to account
for these quark contributions. As will be shown below, the quark
contributions are as significant as the gluon one and increase the
amplitude by about a factor of two. 

The paper is organized as follows. We
explain our approach and consider some examples of diagrams in sect.2. Our
main analytic results are presented in sect.3. The numerical calculations and
comparison with data are given in sect.4.  

\textbf{2}.\hspace{1cm} We use the same approach as in~\cite{Ch0} 
and a very similar notation. So, the presentation will be given in a
short form and we refer to \cite{Ch0} for more detail.

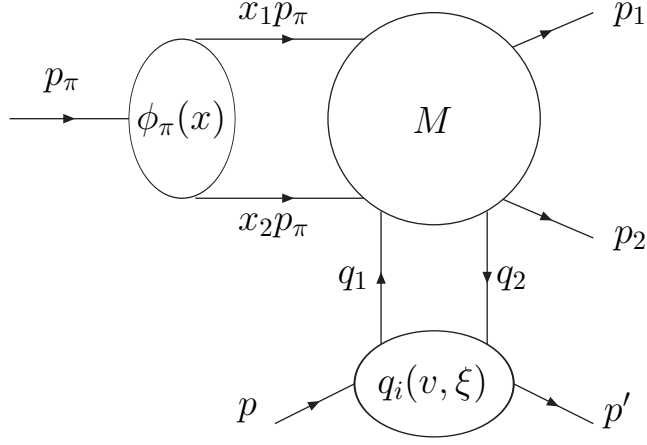
\begin{figure}[ht]
\begin{center}
\begin{picture}(300,180)(0,0)
    \SetColor{Black}
    \ArrowLine(10,125)(55,125)
    \Text(30,140)[]{\large $p_\pi$}
    \COval(75,125)(30,20)(0){Black}{White}
    \Text(75,125)[]{\large $\phi_\pi(x)$}
    \ArrowLine(80,95)(150,95)
    \Text(110,85)[]{\large $x_2 p_\pi$}
    \ArrowLine(80,155)(150,155)
    \Text(110,165)[]{\large $x_1 p_\pi$}
    \ArrowLine(195,95)(230,80)
    \Text(245,80)[]{\large $p_2$}
    \ArrowLine(195,150)(230,165)
    \Text(245,165)[]{\large $p_1$}
    \ArrowLine(150,40)(150,90)
    \Text(140,65)[]{\large $q_1$}
    \ArrowLine(190,90)(190,40)
    \Text(200,65)[]{\large $q_2$}
    \ArrowLine(110,10)(140,25)
    \Text(100,15)[]{\large $p$}
    \ArrowLine(200,25)(230,10)
    \Text(240,15)[]{\large $p'$}
    \COval(170,125)(40,40)(0){Black}{White}
    \Text(170,125)[]{\large $M$}
    \COval(170,25)(20,30)(0){Black}{White}
    \Text(170,25)[]{\large $q_i(v,\xi)$}
\end{picture}
\end{center}
\caption{Kinematics and notations.}
\label{F:kin}
\end{figure}

The diagram in fig.~\ref{F:kin} shows our notation and kinematics of the quark 
contributions to the amplitude. The function $q_i(v,\xi)$ in fig.~\ref{F:kin}
denotes any of the quark distributions of the nucleon:
$i=(u$, $d$, $s$, $\bar{u}$, $\bar{d}$, $\bar{s})$.
The blob $M$ represents the hard kernel of the process.
It is proportional to the scattering amplitude
\begin{equation}
d(x_1 p_{\pi})+\bar{u}(x_2 p_{\pi})+q_i(q_1)\to
d(p_1)+\bar{u}(p_2)+q_i(q_2)
\label{scat}
\end{equation}
of two initial pion quarks on the target quark. In the leading 
twist approximation, all six particles of this process can be considered 
as being on shell. Besides, the pion and target quarks can be considered 
as having zero transverse momenta, as account of primordial virtualities 
and transverse momenta results only in higher twist corrections to $M$. 

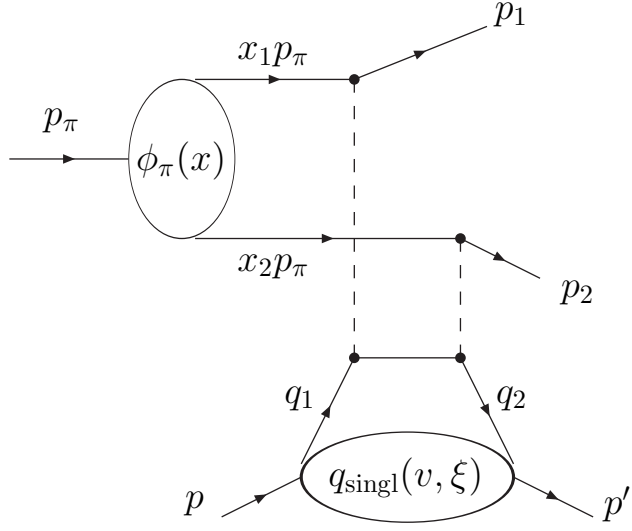
\begin{figure}[ht]
\begin{center}
\begin{picture}(270,200)(0,0)
    \SetColor{Black}
    \ArrowLine(10,155)(55,155)
    \Text(30,170)[]{\large $p_\pi$}
    \COval(75,155)(30,20)(0){Black}{White}
    \Text(75,155)[]{\large $\phi_\pi(x)$}
    \ArrowLine(80,125)(180,125)
    \Text(110,115)[]{\large $x_2 p_\pi$}
    \Vertex(180,125){2}
    \ArrowLine(80,185)(140,185)
    \Text(110,195)[]{\large $x_1 p_\pi$}
    \Vertex(140,185){2}
    \ArrowLine(180,125)(210,110)
    \Text(225,105)[]{\large $p_2$}
    \ArrowLine(140,185)(190,205)
    \Text(200,210)[]{\large $p_1$}
    \DashLine(140,80)(140,185){5}
    \DashLine(180,80)(180,125){5}
    \Vertex(140,80){2}
    \Vertex(180,80){2}
    \Line(140,80)(180,80)
    \ArrowLine(120,40)(140,80)
    \Text(120,65)[]{\large $q_1$}
    \ArrowLine(180,80)(200,40)
    \Text(200,65)[]{\large $q_2$}
    \ArrowLine(90,20)(120,35)
    \Text(80,25)[]{\large $p$}
    \ArrowLine(200,35)(230,20)
    \Text(240,25)[]{\large $p'$}
    \COval(160,35)(17,40)(0){Black}{White}
    \Text(160,35)[]{\large $q_{\text{singl}}(v,\xi)$}
\end{picture}
\end{center}
\caption{The scattering type diagram.}
\label{F:d0}
\end{figure}

In the lowest order in $\alpha_s$, $M$ is given by a set
of connected Born diagrams. Each diagram consists of three quark lines 
connected in all possible ways by exchanges of two hard gluons. There are 
eight such diagrams for each of six $q_i$. As an example, the representatives
of three possible types of diagrams (scattering, annihilation and exchange)
are shown in fig.~\ref{F:d0} and fig.~\ref{F:d1}.

\begin{figure}[p]
\begin{center}
\begin{picture}(270,200)(0,0)
    \SetColor{Black}
    \Text(0,180)[]{\Large\textit{a})}
    \ArrowLine(10,125)(55,125)
    \Text(30,140)[]{\large $p_\pi$}
    \COval(60,125)(30,20)(0){Black}{White}
    \Text(60,125)[]{\large $\phi_\pi(x)$}
    \ArrowLine(65,95)(115,95)
    \ArrowLine(150,95)(200,95)
    \ArrowLine(65,155)(115,155)
    \Line(115,155)(150,155)
    \ArrowLine(150,155)(200,155)
    \Text(90,165)[]{\large $x_1 p_\pi$}
    \Text(90,85)[]{\large $x_2 p_\pi$}
    \Text(205,85)[]{\large $p_2$}
    \Text(205,165)[]{\large $p_1$}
    \DashLine(115,95)(115,155){5}
    \Vertex(115,155){2}
    \ArrowLine(110,40)(115,95)
    \Text(100,65)[]{\large $q_1$}
    \Vertex(115,95){2}
    \DashLine(150,155)(150,95){5}
    \Vertex(150,155){2}
    \ArrowLine(150,95)(155,40)
    \Text(165,65)[]{\large $q_2$}
    \Vertex(150,95){2}
    \ArrowLine(73,10)(103,25)
    \Text(63,15)[]{\large $p$}
    \ArrowLine(163,25)(195,10)
    \Text(205,20)[]{\large $p'$}
    \COval(133,25)(25,35)(0){Black}{White}
    \Text(133,25)[]{\large $q(v,\xi)$}
\end{picture}\\[2cm]
\begin{picture}(270,200)(0,0)
    \SetColor{Black}
    \Text(0,180)[]{\Large\textit{b})}
    \ArrowLine(10,125)(55,125)
    \Text(30,140)[]{\large $p_\pi$}
    \COval(60,125)(30,20)(0){Black}{White}
    \Text(60,125)[]{\large $\phi_\pi(x)$}
    \ArrowLine(65,95)(115,95)
    \ArrowLine(150,95)(200,95)
    \ArrowLine(65,155)(115,155)
    \Line(115,155)(150,155)
    \ArrowLine(150,155)(200,155)
    \Text(90,165)[]{\large $x_1 p_\pi$}
    \Text(90,85)[]{\large $x_2 p_\pi$}
    \Text(205,85)[]{\large $p_2$}
    \Text(205,165)[]{\large $p_1$}
    \DashLine(115,95)(115,155){5}
    \Vertex(115,155){2}
    \ArrowLine(95,20)(150,95)
    \Text(110,65)[]{\large $q_1$}
    \Vertex(115,95){2}
    \DashLine(150,155)(150,95){5}
    \Vertex(150,155){2}
    \ArrowLine(115,95)(170,20)
    \Text(155,65)[]{\large $q_2$}
    \Vertex(150,95){2}
    \ArrowLine(73,10)(103,25)
    \Text(63,15)[]{\large $p$}
    \ArrowLine(163,25)(195,10)
    \Text(205,20)[]{\large $p'$}
    \COval(133,25)(25,35)(0){Black}{White}
    \Text(133,25)[]{\large $q(v,\xi)$}
\end{picture}
\end{center}
\caption{\textit{a}) annihilation type diagram;
\textit{b}) exchange type diagram.}
\label{F:d1}
\end{figure}

The quark and antiquark distributions in the nucleon are defined as
($\hat{P}=\gamma_{\mu}P_{\mu}$):
\begin{equation}
\langle P'|\bar{q}(0)_{\beta}\,q(w)_{\alpha}|P \rangle =
\frac{2\hat{\bar{P}}_{\alpha \beta}}{4}\int_{-\xi}^{1}dv\,
\left\{ q(v,\xi)\,e^{-i(v+\xi)(\bar{P}w)}-\bar{q}(v,\xi)\,  
e^{i(v-\xi)(\bar{P}w)}\right\},
\label{dist}
\end{equation}
where $q=u$, $d$, $s$, while the kinematical variables are:
\begin{eqnarray}
&&
q_1=(v+\xi)\bar{P},\quad
q_2=(v-\xi)\bar{P},\quad
\bar{P}=(P+P')/2\,,
\nonumber\\&&
\Delta=(q_1-q_2)=2\xi\bar{P},\quad
\xi=\frac{\p}{2 y_1 y_2 s},\quad
z_1=\frac{v+\xi}{2\xi},\quad 
z_2=\frac{v-\xi}{2\xi},
\nonumber\\&&
z_1-z_2=1,\quad
2\,(p_{\pi}\Delta)=M^2=\frac{\p}{y_1 y_2},
\label{kin}\\&&
p_1=y_1 p_{\pi}+y_2 \Delta+k_{\perp}+(q_{\perp}/2),\quad
p_2=y_2 p_{\pi}+y_1 \Delta-k_{\perp}+(q_{\perp}/2),
\nonumber
\end{eqnarray}
where $q_{\perp}$ is the small final transverse momentum of the nucleon while
$k_{\perp}$ is large.

The structure of the amplitude is (symbolically):
\begin{equation}
T \sim \langle P'|\bar{q}\cdot q|P \rangle \otimes (\bar{\psi}_1 M \psi_2)
\otimes \langle 0|\bar{u}\cdot d|\pi^{-} \rangle,
\label{symb}
\end{equation}
where the first matrix element introduces the skewed quark and antiquark
distributions of the nucleon, $q_j(v,\xi)$ and $\bar{q}_j (v,\xi)$, 
$\bar{\psi}_1$ and $\psi_2$ are the free spinors of final 
quarks, $M$ is the hard kernel, i.e.\ the product of all vertices and hard 
propagators, the last matrix element introduces the pion wave function
$\phi_{\pi}(x)$, and $\otimes$ means the appropriate convolution. 

As an example let us consider the diagram in fig.~\ref{F:d0} as it gives the 
main contribution. Proceeding in the above described way (see also~\cite{Ch0}),
one obtains for the sum of the diagram in fig.~\ref{F:d0} and the same diagram but
with crossed gluon lines (the Feynman gauge is used for gluons):
\begin{eqnarray}
T_0 &{}={}& \frac{\rho_0}{y_1 y_2}\int_{0}^{1}\frac{dx_1\,\phi_{\pi}(x_1)}{x_1 x_2}
\left\{ \int_{-\xi}^{1}\frac{dv\,q_{\text{singl}}(v,\xi)\,
[x_1-y_1-(v/\xi)]}{[v(x_1-y_1)-\xi(x_1 y_2+x_2 y_1)]}+(1\leftrightarrow 2)
\right\},\nonumber\\
\rho_0 &{}={}& \delta_{kl}
\frac{(4\pi\alpha_s)^2 f_{\pi}}{27}
(\bar{\psi}_1 \hat{\Delta} \gamma_5 \psi_2)\,
\frac{(y_1 y_2)^2}{\mathbf{k}_{\perp}^4};\quad
(1\leftrightarrow 2)\equiv \{ x_1\leftrightarrow x_2;\,y_1\leftrightarrow y_2\},
\label{T0}
\end{eqnarray}
where $k$, $l$ are the colour indices of final quarks.
As it is expected that the imaginary part of the amplitude is the main one 
at high energies, we show explicitly below only its value. From eq.~(\ref{T0}) one
obtains\footnote{We introduce the terms $i\epsilon$ into propagators through
$s+i\epsilon$, i.e.\ $\xi\to\xi-i\epsilon$ in all diagrams.
Besides, it is implied that the quark distributions $q_j(v,\xi)$
have simple zero at $v=-\xi$.}:
\begin{eqnarray}
\Im T_0 &{}={}& \frac{\pi\rho_0}{y_1 y_2}
\int_0^1\frac{dx_1\,\phi_{\pi}(x_1)\,(x_1 x_2+y_1 y_2)\,
q_{\text{singl}}(|\bar{v}|,\xi)}{x_1 x_2 (x_1-y_1)^2}\,
\Theta(|x_1-y_1|>\delta),
\nonumber\\
q_{\text{singl}} &=& \sum_{j=u,d,s} (q_j+\bar{q}_{j})\,,\quad
\bar{v} = \xi\,\frac{x_1 y_2+x_2 y_1}{x_1-y_1}\,,\quad
\delta = \frac{\p}{s}\,.
\label{ImT0}
\end{eqnarray}

As for all other diagrams, there are simple relations between contributions 
to the hard kernel M of different quark distributions for these diagrams:
\begin{eqnarray}
M_{\bar{d}}(1,2,v) &{}={}& - M_{d}(1,2,-v);\quad
M_{\bar{u}}(1,2,v) = M_{d}(2,1,v);
\nonumber\\
M_{u}(1,2,v) &{}={}& - M_{d}(2,1,-v).
\label{symm}
\end{eqnarray}

\textbf{3.}\hspace{1cm} Summing up the contributions of all diagrams, one
obtains for the total amplitude:
\begin{equation}
T_{\text{total}} = (\,T_{\text{gluon}}+T_0+T_u+T_d+T_{\bar{u}}+
T_{\bar{d}}\,)\,,
\label{Ttot}
\end{equation}
where $T_{\text{gluon}}$ is the contribution of the gluon distribution
(see~\cite{Ch0}, eqs.~(12--17)), $T_0$ is given by eqs.~(\ref{T0}--\ref{ImT0}),
while the rest terms in eq.~(\ref{Ttot}) are as follows:
\begin{eqnarray}
\Im T_j &{}={}& \frac{8\pi\rho_0}{3} \int_0^1 dx_1\,\phi_{\pi}(x_1)\,\Sigma_j\,,
\nonumber\\
\Sigma_u &{}={}& \frac{q_u(\xi,\xi)}{x_1 y_2}
+\frac{[x_1^2 x_2-(x_1-y_1)^3]\,q_u(|\bar{v}|,\xi)\,\Theta_{+}}
{8\,|x_1-y_1|\,x_1^2 x_2 y_1 y_2^2}
+\frac{q_u(|\bar v|,\xi)\,\Theta_{-}}{8\,|x_1-y_1|\,x_1 x_2^2},
\label{Tu}
\end{eqnarray}
where $\Theta_{+}=\Theta(y_1+\delta <x_1 < 1)$,
$\Theta_{-}=\Theta (0< x_1<y_1-\delta)$;
\begin{eqnarray}
\Sigma_d &{}={}& \frac{(1+y_1)\,q_d(\xi,\xi)}{x_1 y_1^2}
\nonumber\\
&&{} + \frac{(1-2\,x_1 x_2)\,y_2\,q_d(\xi,\xi)}{8\,x_1 x_2^2 y_1^2}
- \frac{\left[x_1 x_2^2 + (x_1-y_1)^3\right]\,q_d(|\bar{v}|,\xi)\,\Theta_+}
{8\,|x_1-y_1|\,x_1 x_2^2 y_1^2 y_2}
\nonumber\\
&&{} + \frac{(1-2\,x_1 x_2)\,q_d(\xi,\xi)}{8\,x_1^2 x_2 y_1}
- \frac{q_{d}(|\bar{v}|,\xi)\,\Theta_{-}}{8\,|x_1-y_1|\,x_1^2 x_2}\,.
\label{Td}\\
\Sigma_{\bar{d}}&(1,&2) = \Sigma_u(2,1;\,q_u\to q_{\bar{d}})\,;\quad
\Sigma_{\bar{u}}(1,2) = \Sigma_d(2,1;\,q_d\to q_{\bar{u}})\,.
\label{Tother}
\end{eqnarray}

Let us note that while the separate terms
in $\int dx\,\phi_{\pi}(x)\,\Sigma_j$ are logarithmically divergent
at $x_{1,2}\to 0$, it is not difficult 
to see that the divergences cancel in the sum, so that the integral is 
finite. This is an important point, as it shows that the whole approach is
self-consistent, i.e. the hard kernel remains hard and the soft end point 
regions $x_{1,2}\to 0$ give only power suppressed corrections. 

Let us point out also 
that, unlike the gluon contribution, the quark contribution to the amplitude 
is not symmetric at $y_1\leftrightarrow y_2$. However, this asymmetry is 
very small numerically in the main part of the phase space and becomes
significant at the very edges $y_{1,2}\to 0$ only, where the
cross section is very small by itself.

The expressions~(\ref{T0}--\ref{Tother})
constitute the main result of this paper. 

\textbf{4.}\hspace{1cm} We present in this section the numerical estimates of
the total cross section with account of both the gluon and quark 
contributions.

The quark distributions in the nucleon, $q_j(v,\xi,t=0,\mu)$ and
$\bar{q}_j(v,\xi,t=0,\mu)$ at $\mu\simeq k_{\perp}=2$\,GeV, are chosen in a
simple form as (at $v\geq\xi$ where we only need them):
\begin{eqnarray}
q_u(v,\xi) &{}={}& 4\,v^{-0.25}\,(1-v)^3 + q_{\text{nonval}}(v,\xi)\,,
\nonumber\\
q_d(v,\xi) &{}={}& 2\,v^{-0.25}\,(1-v)^3 + q_{\text{nonval}}(v,\xi)\,,
\label{qmod}\\
\bar{q}_u = \bar{q}_d &{}={}& q_s = \bar{q}_s = q_{\text{nonval}}(v,\xi)
= 0.13\,v^{-1.2}\,(1-v)^7.
\nonumber
\end{eqnarray}
These forms agree reasonably well with the results of numerical calculations 
in~\cite{GRV} and~\cite{Rys}. The nuclear form factor is introduced
as in~\cite{Ch0}, and we consider for comparison the same two model forms of the 
pion wave function: 
\begin{equation}
\phi_{\pi}^{\text{asy}} = 6\,x_1 x_2\,;\quad
\phi_{\pi}^{\text{CZ}}(x,\,\mu\simeq 2\,\text{GeV})
= 15\,x_1 x_2 \Bigl[(x_1-x_2)^2+0.2\Bigr].
\label{wf}
\end{equation}

Because the quark contributions have some asymmetry between the mirror 
points at $y_1<0.5$ and $y_1>0.5$, while the experiment does not 
distinguish between the $d$ and $\bar{u}$ jets, we give below the
answers for the cross sections symmetrized in $y_1\leftrightarrow y_2$
(i.e.\ the half sum of values at the mirror points).

To show the role of quark contributions we present in figs.4a and 4b
the cross sections calculated with pure gluon distributions and the total 
cross sections (i.e.\ with account of both the gluon and quark distributions),
for $\phi_{\pi}^{\text{CZ}}$ and $\phi_{\pi}^{\text{asy}}$ separately.
It is seen that the cross section increases roughly 3 times 
with account of quark contributions. Let us illustrate in more detail how
this happens for the case $\phi_{\pi}^{\text{CZ}}$
at the middle point $y_1=y_2=0.5$. For the gluon distribution, the main
contribution comes from the diagram in fig.~3 in~\cite{Ch0} (the gluon analog
of the diagram in fig.~\ref{F:d0} in this paper). It contributes (36.2) to the 
amplitude in some units, while all other diagrams with the gluon distribution 
contribute $(-19.3)$. So, the total gluon contribution becomes (16.9). The 
quark diagram in fig.~\ref{F:d0} contributes (9.4), 
while all other quark diagrams 
contribute (2.9). So, the total quark contribution is (12.3), and it is 
quite comparable with the total gluon contribution. When moving from 
the middle point to the edges $y_{1,2}\to 0$, the role of quark
contributions only increases.
The picture for $\phi_{\pi}^{\text{asy}}$ is very similar.

\begin{figure}[p]
\begin{center}
\begin{picture}(456,552)
\put(228,414){\makebox(0,0){\includegraphics[width=.63\textwidth]{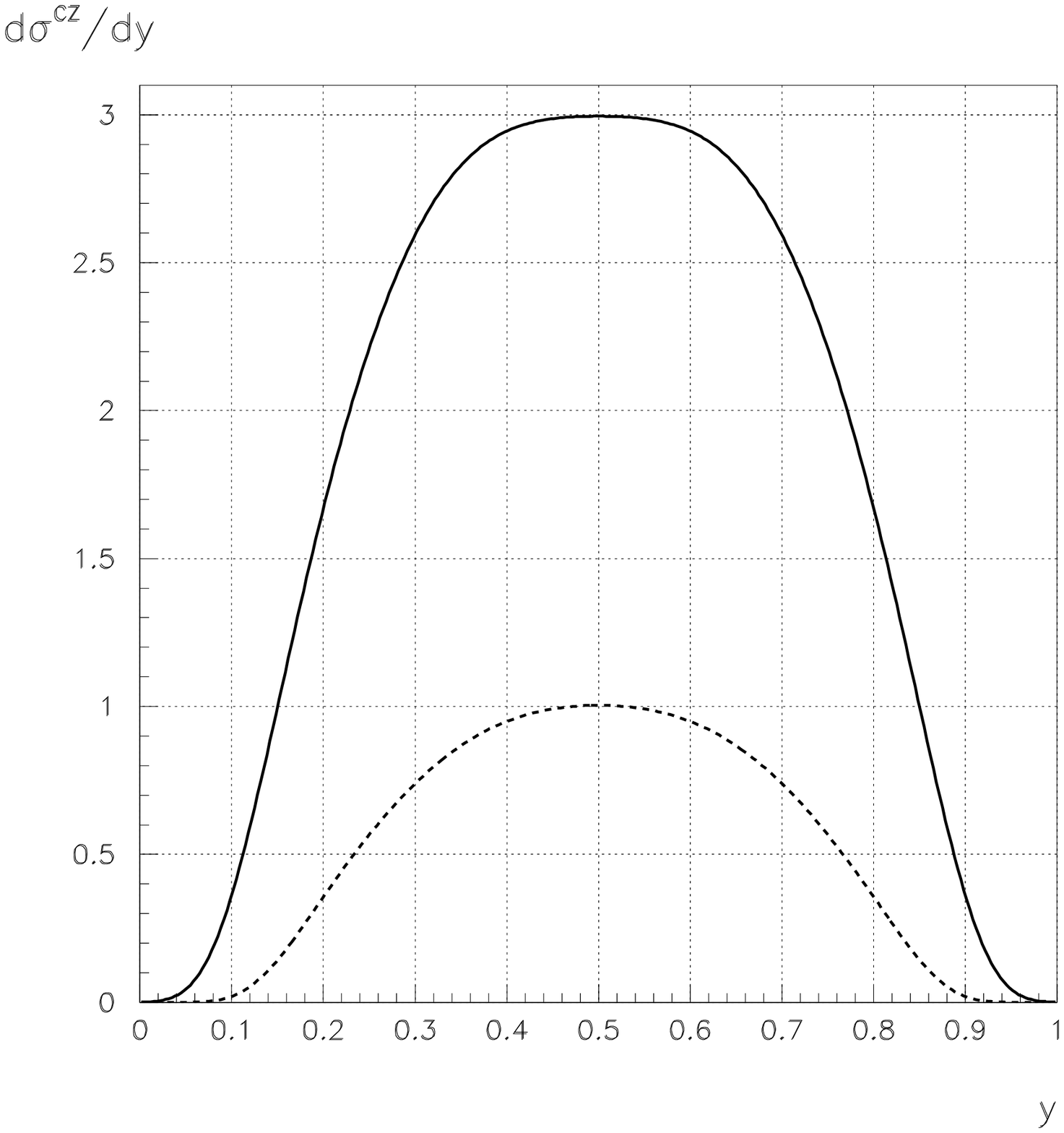}}}
\put(228,138){\makebox(0,0){\includegraphics[width=.63\textwidth]{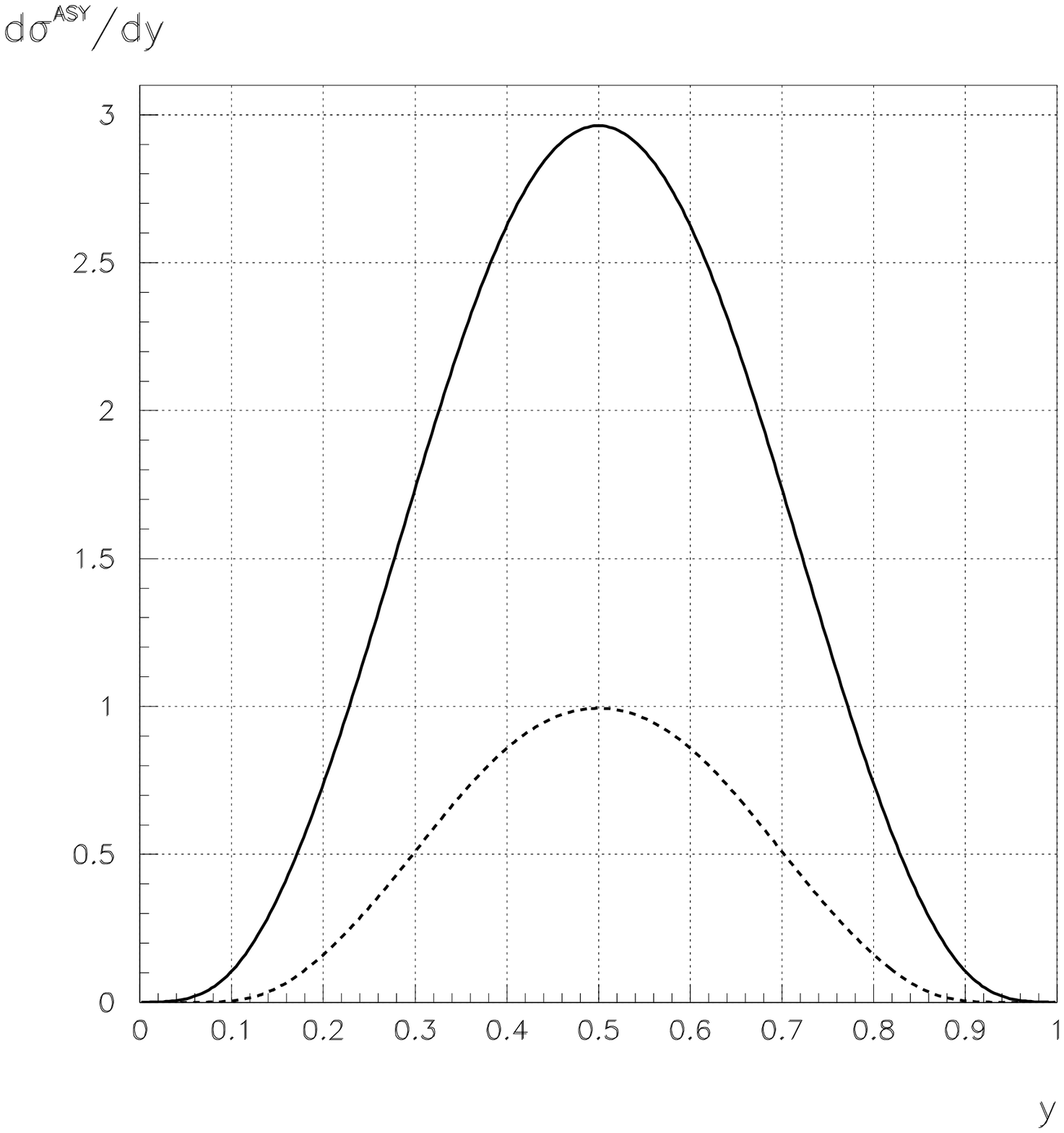}}}
\put(70,514){\Large\textit{a})}
\put(70,238){\Large\textit{b})}
\end{picture}
\end{center}
\caption{The relative values of the pure gluon (dashed line) and total
(quark+gluon, solid line) cross sections
for \textit{a}) $\phi_\pi^{\text{CZ}}(x)$
and \textit{b}) $\phi_\pi^{\text{asy}}(x)$.}
\label{F:qg}
\end{figure}

\begin{figure}[p]
\begin{center}
\includegraphics[width=\textwidth]{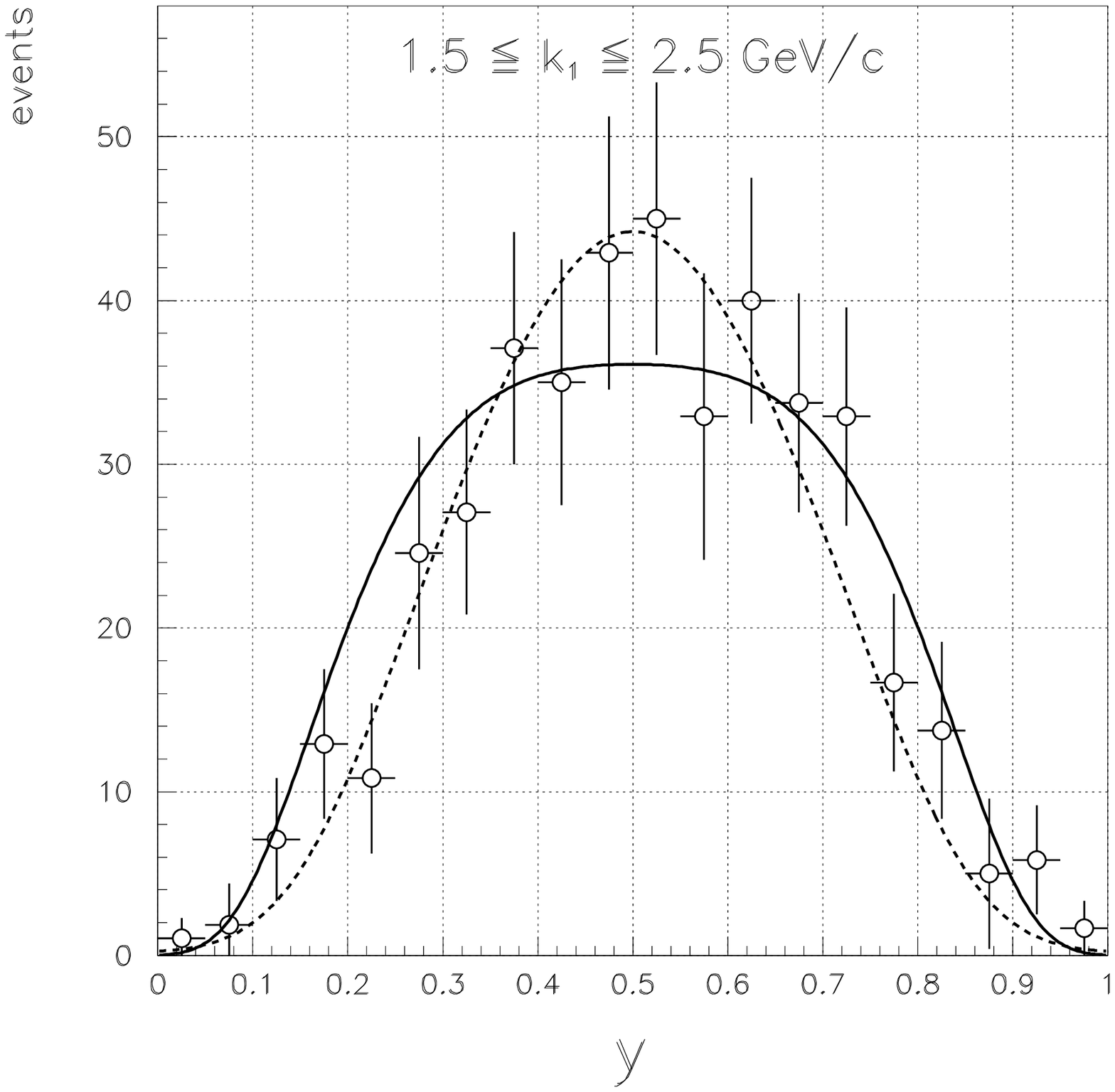}
\end{center}
\caption{The y-distribution of jets for the total (quark+gluon) cross
section, calculated for $k_{\perp}=2$\,GeV,
$E_{\pi}=500$\,GeV and with the pion wave functions:
$\phi_{\pi}^{\text{CZ}}(x,\mu\simeq 2\,GeV)$ --- solid line,
$\phi_{\pi}^{\text{asy}}(x)$ --- dashed line.
The overall normalization is arbitrary, but the relative normalization of 
two curves is as calculated. The data points are from the E791
experiment~\cite{Aitala}.}
\label{F:exp}
\end{figure}

We show in fig.~\ref{F:exp} the calculated total cross sections in comparison with 
the data from the E791 experiment~\cite{Aitala}. Comparing these with the 
gluon contributions (see fig.~5 in~\cite{Ch0}) one sees that the
difference between $\phi_{\pi}^{\text{CZ}}$ and $\phi_{\pi}^{\text{asy}}$
becames slightly
more pronounced, but the main characteristic properties of the distribution
of jets in $y$ remained the same. I.e., the cross section is not much 
sensitive to the profile of the pion wave function $\phi_{\pi}(x)$. It
seems that the present experimental accuracy is insufficient for 
obtaining stringent enough restrictions on the form of the pion 
wave function from these data.

Besides the experimental accuracy, it will
be desirable, in addition, to improve the theoretical accuracy:
to find one loop corrections to the hard kernels,
to estimate higher twist corrections,
to account for the real part of the amplitude, etc.
Moreover, the main contributions from the diagram in fig.~3 in~\cite{Ch0}
and the diagram in fig.~\ref{F:d0} here
are sensitive to the precise form of the gluon and quark distributions in
the nucleon, which are considered as known in our approach. So, 
the dependence of the $y$-distribution of jets on allowed variations of 
the forms of $g(v,\xi)$ and $q_i(v,\xi)$ 
has to be considered in more detail also.

Finally, let us give some typical numbers for the absolute values of the cross
sections. For $E_{\pi}=500$\,GeV, $k_{\perp}=2$\,GeV and $\phi_{\pi}^{\text{CZ}}$: 
$d\sigma_{\text{Pt}}^{\text{CZ}}/(dk_{\perp}^2 dy) \simeq
7.5\cdot 10^{-2}\,\text{mbarn}\cdot\text{GeV}^{-2}$
at the middle point $y_1=y_2=0.5$. The ratio $d\sigma^{\text{asy}}/
d\sigma^{\text{CZ}}$ is $\simeq 1.23$ at $y_1=0.5$ and $\simeq 0.68$ at $y_1=0.25$.

\end{document}